\documentclass[3p]{elsarticle}
\usepackage{lineno,hyperref}
\usepackage{graphicx}
\usepackage{amssymb}
\usepackage{amsmath}
\usepackage{adjustbox}
\modulolinenumbers[5]

\journal{Journal of \LaTeX\ Templates}

\begin{document}

\begin{frontmatter}

\title{Formation of the stopped polarization pulse in a rectangular quantum well}

\author{P.~A.~Belov$^{1,2}$}
\ead{pavelbelov@gmail.com}
\address{$^{1}$Institut f\"{u}r Physik, Universit\"{a}t Rostock, Albert-Einstein-Strasse 23, 18059 Rostock, Germany}
\address{$^{2}$Spin Optics Laboratory, St. Petersburg State University, Ulyanovskaya 1, 198504 St. Petersburg, Russia}
\author{R.~M.~Arkhipov$^{3}$}
\address{$^{3}$St. Petersburg State University, Universitetskaya nab. 7, 199034 St. Petersburg, Russia}
\begin{abstract}
The induced polarization oscillations in a one-dimensional rectangular quantum well are modeled by a numerical solution of the time-dependent Schr\"{o}dinger equation. The finite-difference discretization over time is realized in the framework of the Crank-Nicolson algorithm, whereas over the spatial coordinate it is combined with the exterior complex-scaling technique. A formation of the harmonic oscillations of the dipole moment by an incident short unipolar pulse is shown. It is obtained that the frequency of oscillations is solely defined by the energy of the main resonant transition. Moreover, if two such short unipolar pulses are delayed by a half-period of the oscillation, then these oscillations can be abruptly induced and stopped. Thus, the so-called stopped polarization pulse is obtained. It is shown that both the amplitude and the duration of the incident unipolar pulse, contributing to the so-called electric pulse area, define the impact of the incident pulse on the quantum system.
\end{abstract}
\begin{keyword}
polarization, dipole moment, quantum well, Schr\"{o}dinger equation, complex scaling, Crank-Nicolson method
\end{keyword}
\end{frontmatter}

\section{Introduction}

The technology developments related to a generation of the ultrashort electromagnetic pulses are motivated by the use of such pulses to study the dynamics of ultrafast processes on atomic time scales~\cite{Calegari,Hassan,Garg,Biegert}.
To date, the pulses with a duration in the femtosecond and attosecond ranges have been experimentally obtained by several groups~\cite{Mourou,Khazanov,Krausz,Midorikawa,Goul}.
The limit for reducing the duration of electromagnetic pulses is to obtain a single-cycle pulse containing two half-waves of the electric field of opposite polarity or a unipolar pulse in which the field does not change sign during the duration of the pulse. An interest in the unipolar pulses is related to the enhanced ultrafast control of quantum systems, in comparison to bipolar multicycle pulses~\cite{eight}.

Recently, the possibility for obtaining single-cycle and unipolar subcycle pulses due to collective spontaneous emission (superradiance) of the so-called stopped polarization pulse (SPP) has been proposed~\cite{Arkhipov1,Arkhipov2,Arkhipov3,Arkhipov4,Arkhipov5,Arkhipov6,eleven}. These results are summarized in the reviews~\cite{nine,ten}. The idea of the approach is as follows. An extended medium is excited by a pair of ultrashort pump pulses following with a delay of a half-period of a main resonant transition of the medium. The first pump pulse excites polarization oscillations of the medium, while the second one stops them. For such a delay, a polarization half-wave, the so-called SPP, arises. It is a source of radiation of unipolar or single-cycle pulses in the optical or terahertz frequency range.
This mechanism can underlie the generation of unipolar rectangular or triangular pulses in an inhomogeneous Raman active medium (RAM)~\cite{twelve}.

In the studies listed above, the medium was modeled using the two-level or three-level quantum systems~\cite{nine,ten}. As well, the model of a classical non-linear harmonic oscillator was employed to simulate the nonlinear medium such as the RAM response.
In a model of the classical oscillator or a nonlinear radiating oscillator, the problem of its excitation and subsequent de-excitation looks rather trivial~\cite{Arkhipov1}. When a charged particle is excited by an extremely short pulse, much shorter than the period of oscillations, it is displaced from the equilibrium position, moves to the position of maximum displacement from the equilibrium, and then, at the end of the first half-period of oscillations, again returns to the equilibrium position.
If there is no dissipation, the absolute value of the momentum after a half-period of oscillations is equal to that acquired from the first pulse. However, it has the opposite direction.
To stop the motion of the oscillator, the second pulse, identical to the first one, should be applied.

In an extended medium, two such short pulses will create an unusual object, namely, a moving half-wave of macroscopic polarization of the medium, which will be a source of radiation contained within this half-wave of polarization between two pulses~\cite{eleven,twelve}.
However, this simple consideration is possible only when the medium is described in the approximation of linear and nonlinear oscillators or in the framework of a two-level model.
In fact, for modeling the excitation of quantum systems (atoms, molecules, etc.) by extremely short pulses, one should take into account larger number of energy levels.

The real medium has many energy levels. When several levels are excited by a short pulse, the system switches to a superposition state and the resulting dipole moment oscillates at the frequencies of allowed transitions. Therefore, not a single oscillator with the only frequency, as in the classical case, but several ``oscillators'' with their own frequencies simultaneously appear in the system. They create their own polarization waves.
The oscillations of the dipole moment keep after the end of action of the excitation pulse. Generally speaking, they continue during the polarization relaxation time, usually denoted as $T_{2}$~\cite{Boyd}. The value of $T_{2}$, for example for quantum dots, can vary over a wide range: from hundreds of femtoseconds to a few picoseconds at room temperature and reach hundreds of nanoseconds at cryogenic temperatures~\cite{timeT2}.
Therefore, for the medium characterized by a relatively large dipole relaxation time, the oscillations last much longer than the period of oscillations.
In such a situation, the second pulse, if it is identical to the first one and its arrival time is a half-period of oscillations between the ground and the first excited state, may not be able to stop the movement of other ``oscillators''. As a result, the polarization of the medium is not damp.
Therefore, a previously unconsidered question about a possibility of generating the SPP in the multilevel medium naturally arises.

In this paper, we model the conditions for an emergence of the SPP in the one-dimensional rectangular quantum well (QW) with finite barriers using a numerical solution of the time-dependent Schr\"{o}dinger equation (TDSE).
The heights of the barriers are chosen to simulate the required number of bound states in QW.
The TDSE does not model dissipation, therefore the dipole relaxation time is assumed to be infinite.
In order to numerically solve the TDSE, we employ the finite-difference discretization~\cite{Khramtsov,Belov2017,BelovPhysE,Loginov} over time and space variables. The time evolution is realized via the absolutely stable Crank-Nicolson scheme~\cite{CN}. Over the spatial variable, the exterior complex-scaling method was employed to avoid the reflections from the rigid-wall boundary conditions at the asymptotics~\cite{Moiseyev,Rescigno,MoiseyevTDSE,Zavin,Vinitsky}.
The oscillations of the polarization were numerically studied as a function of the incident Gaussian pulse width and the conditions for emergence of the SPP were determined.
It was shown that such conditions are related to the so-called electric pulse area~\cite{SE01,SE02}, defining the effect of the incident pulse on a quantum system.

\section{Theory}
The interaction of a charged particle, located in a finite-depth QW, with an intense external field is described by the TDSE for the electron wave function $\Psi(x,t)$. Assuming a one-dimensional geometry of the studied system and the dipole approximation (the electric field does not depend on the coordinate), in the length gauge the TDSE reads as~\cite{int02,Vinitsky,Rosanov2018}
\begin{equation}
i\hbar \, \frac{\partial \Psi(x,t)}{\partial t} = \Big[ -\frac{\hbar^{2}}{2m} \, \frac{\partial^{2}}{\partial x^{2}} + V(x) -e\,x \, E(t) \Big] \Psi(x,t),
\label{SE}
\end{equation}
where the rectangular QW potential is given as
\begin{equation}
\label{U0}
V(x) = \left\{
  \begin{array}{lr}
    0 & \mbox{ if  } |x| < a/2 \\
    V & \mbox{ if  } |x| \ge a/2
  \end{array}
\right. .
\end{equation}
Here $V$ and $a$ are the barrier height and the QW width, respectively. These parameters were chosen to simulate the required number of bound states in QW. Other quantities are following: $m$ is the electron mass, $e$ is the charge, $\hbar$ is the reduced Planck constant.

For the stationary quantum systems, the wave function of the $m$-th eigenstate is presented as $\Psi_{m}(x,t)=\Psi_{m}(x) e^{-iE_{m}t/\hbar}$.
The evolution of an arbitrary state can be written as a superposition of eigenstates with coefficients $a_{m}(t)$.
As a result, the polarization (the dipole moment)~\cite{Landau}
$$
P(t)=
\text{Re} \left[ a^{\star}_{m}(t) a_{n}(t) e^{iw_{mn}t} \int \Psi^{\star}_{m}(x) \,
(e\, x)\, \Psi_{n}(x) dx \right]
$$
oscillates with a frequency determined by energies $w_{mn}=(E_{m}-E_{n})/\hbar$ of the resonant transitions.

In our study, the quantum system is excited by the two incident unipolar pulses which can be presented in the form
\begin{equation}
\label{Et}
E(t) = E_{0} \exp{\left(-t^2/\tau_p^2 \right)} + E_{0} \exp{\left(-(t-\tau_{d})^2/\tau_p^2 \right)}, 
\end{equation}
where $E_{0}$ is the pulse amplitude, $\tau_{d}$ is the delay between pulses, and $\tau_p$ is the relatively small pulse width.

We numerically model how the polarization of the multilevel system, excited by the pair of the incident short unipolar pulses, oscillates with time.
To this end, we consider a QW having five bound states and excite it by the two short Gaussian pulses~(\ref{Et}) delayed by half-period $\tau_{d}=\pi/w_{21}$ of the resonant transition from the second level to the first one.
We assume that the dipole moment oscillations last undistorted much longer than $\tau_{d}$, i.e. the dipole relaxation time, $T_{2}$, is taken to be infinite~\cite{Boyd}.
An observation of the resulted time-evolution of the polarization for different widths of the incident Gaussian pulses allows us to draw a conclusion about the effect of such pulses and the feasibility to obtain the SPP.

\section{Numerical modeling}

The modeling is based on the numerical solution of the TDSE.
For such problems, the Dirichlet problem with the trivial boundary conditions is usually introduced over the space variable, whereas over time the Cauchy problem with the initial condition is established. 
The trivial (or rigid-wall) boundary conditions lead to reflections of the outgoing waves, a quantization of the continuum, and, thus,
do not allow modeling of ionization effects in open quantum systems.
To overcome this issue, i.e. to correctly model an open system, the artificial negative imaginary potentials, the so-called absorbing potentials, are usually employed~\cite{abs1,abs2,abs3}.
Another possible way is to use the integral boundary conditions~\cite{int02,int1}.
However, for the time-dependent problems the application of such conditions is difficult because they require integrating the solution over all previous moments of time~\cite{int2}.
To avoid such an integration, different simplified and approximate absorbing boundary conditions were used~\cite{abs0,int01}.

In order to model the nonresonantly excited open quantum system, we applied the so-called complex-scaling technique which is free from the mentioned computational shortcomings.
For stationary problems, this technique has been established and proved in Refs.~\cite{Aguilar,Balslev,Simon,SimonECS}.
Since then, many stationary scattering problems in nuclear and atomic physics have been studied by the complex scaling~\cite{Weinhold,Ho,Kukulin,Moiseyev,Elander,Zavin,Vinitsky,Telnov,yakovlev,Myo,belovFB,Zaytsev,BelovPRB}.
It states that the complex energies of resonant states can be identified in the discrete spectrum of the non-Hermitian Hamiltonian $H_{\theta}=H(x \, \text{exp}(i\theta),y,z)$, obtained from the initial one $H(x,y,z)$ by the scaling of the coordinate $x$ by a complex factor $\text{exp}(i\theta)$.
Here $\theta>0$ ($\theta<0$) is the scaling angle, i.e. the angle of a rotation of the coordinate $x>0$ ($x<0$) into the upper (lower) complex half-plane.
Such a scaling allows one to associate resonant states
with the discrete spectrum by making the outgoing scattering waves
to be square-integrable by an appropriate choice of $\theta$.
\begin{figure}[th!]
\begin{center}
\includegraphics[angle=0, width=0.5\linewidth]{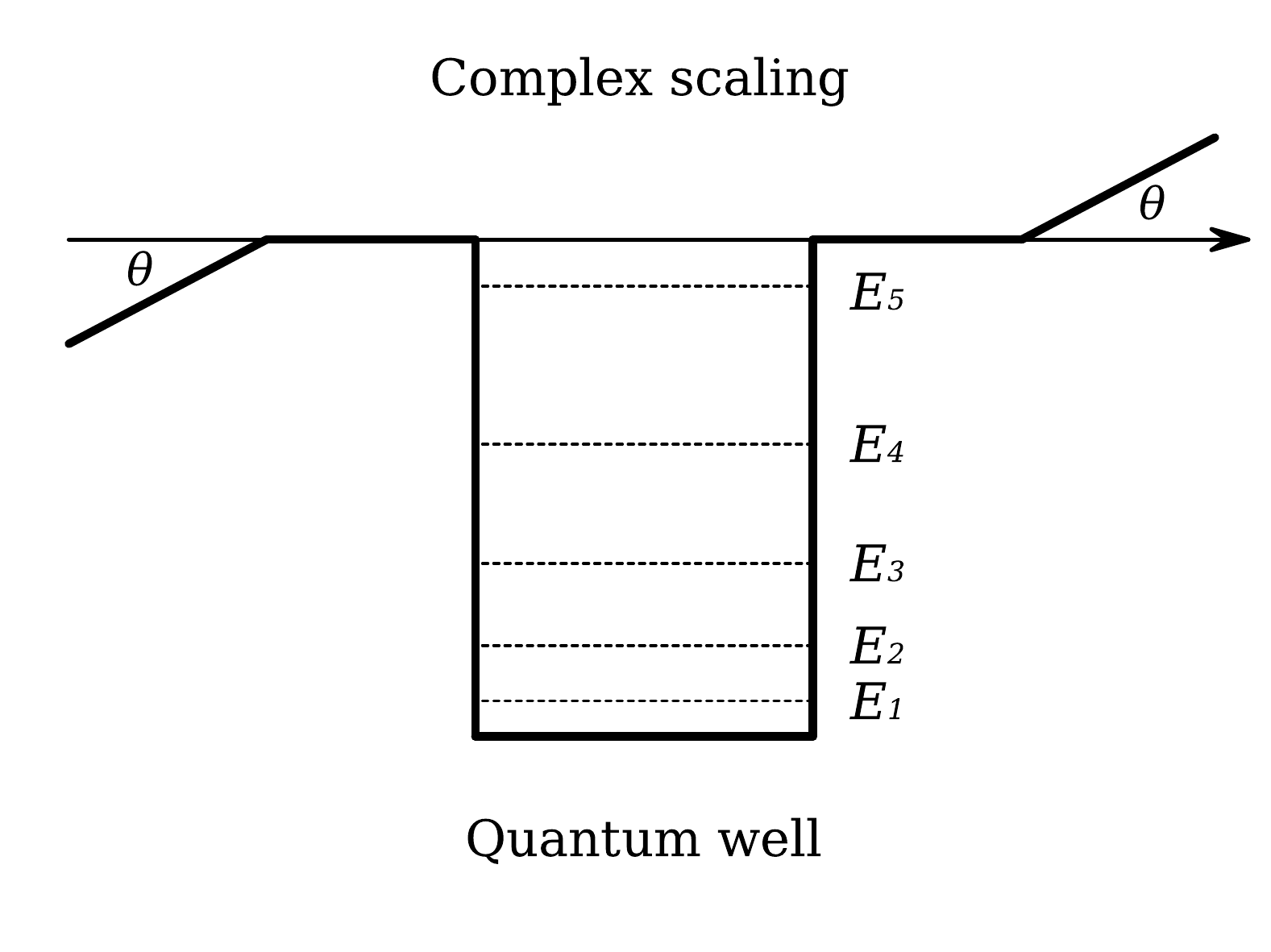}
\caption{The exterior complex-scaling rotation applied to the one-dimensional quantum well potential with five bound states. The angle of the rotation is explicitly denoted as $\theta$.}
\label{qwcs}
\end{center}
\end{figure}

The complex scaling is also appropriate for studying the resonances in time-dependent problems of interaction of the quantum system with an incident electromagnetic field.
It is shown in Ref.~\cite{MoiseyevTDSE} that asymptotically, as $|x|\to \infty$, the interaction with the electromagnetic field vanishes and a free motion takes place there.
As a result, for the problem~(\ref{SE}) with the rectangular QW potential defined by Eq.~(\ref{U0}) we can apply the so-called exterior complex-scaling where the rotation is performed in the asymptotic domain.
It introduces the change of the variable $x \in [-X/2,X/2]$ according to the formula:
\begin{equation}
\label{Rx}
R(x) = \left\{
  \begin{array}{lr}
    x & \mbox{ if  } |x| \le x_{CS} < X/2 \\
    x_{CS} + (x - x_{CS}) \exp{(i\theta)} & \mbox{ if } x_{CS} < x \le X/2 \\
    -x_{CS} - (|x| - x_{CS}) \exp{(i\theta)} & \mbox{ if } -X/2 \le x < -x_{CS}
  \end{array}
\right.
\end{equation}
where the parameter $x_{CS}>a$ defines the asymptotic domain.
This exterior complex-scaling rotation is shown in Fig.~\ref{qwcs}. On the one hand, such a complex scaling damps the outgoing waves outside the QW by generating the exponentially vanishing functions. On the other hand, it keeps the interaction region unperturbed and, thus, does not change the dynamics of the quantum system.

In general, the rotation $R=R(x)$ leads to the change of the second derivative over $x$ in Eq.~(\ref{SE}) which is given as
\begin{equation}
\label{rot}
\frac{\partial^{2}}{\partial R^{2}} = \frac{1}{(R'_{x})^{2}} \frac{\partial^{2}}{\partial x^{2}} - \frac{R''_{xx}}{(R'_{x})^{3}} \frac{\partial}{\partial x},
\end{equation}
where $R'_{x}$ and $R''_{xx}$ are derivatives of $R(x)$,
depending on the particular choice of $R(x)$.
In our case, to keep continuity of these partial derivatives we smoothed the cusps of the rotation~(\ref{Rx}) by appropriate smooth functions as it was done in Ref.~\cite{BelovPRB}.

The unknown quantities, i.e. the polarization (the dipole moment) and populations of energy levels, as functions of time are obtained by the numerical integration over coordinate of the corresponding operator~$\mathcal O$
\begin{equation}
\label{DM}
\langle \Psi |\mathcal O| \Psi \rangle = \int_{-\infty}^{\infty} {\mathcal O}(x) |\Psi(x,t)|^{2} \, dx \approx \int_{C} \tilde{\mathcal O}(R(x)) |\Psi(R(x),t)|^{2} \, dR(x),
\end{equation}
where $\tilde{\mathcal O}(R(x))$ is the numerically obtained complex-scaled operator and the integration is performed over the contour $C$ defined by $R(x)$.

The simulation of the excitation of energy levels in the rectangular QW was performed using the numerical solution of the TDSE~(\ref{SE}), taking into account the complex rotation of the coordinate variable~(\ref{Rx}). The discretization of the equation was carried out by the finite-difference method~\cite{Khramtsov,Belov2017,BelovPhysE,Loginov}. The complex-scaled wave function of the ground state was the initial condition for the time evolution.
The numerical scheme and a comparison of the benchmark results are given in the Appendix.
%
\begin{figure}[th!]
\begin{center}
\includegraphics[angle=0, width=0.5\linewidth]{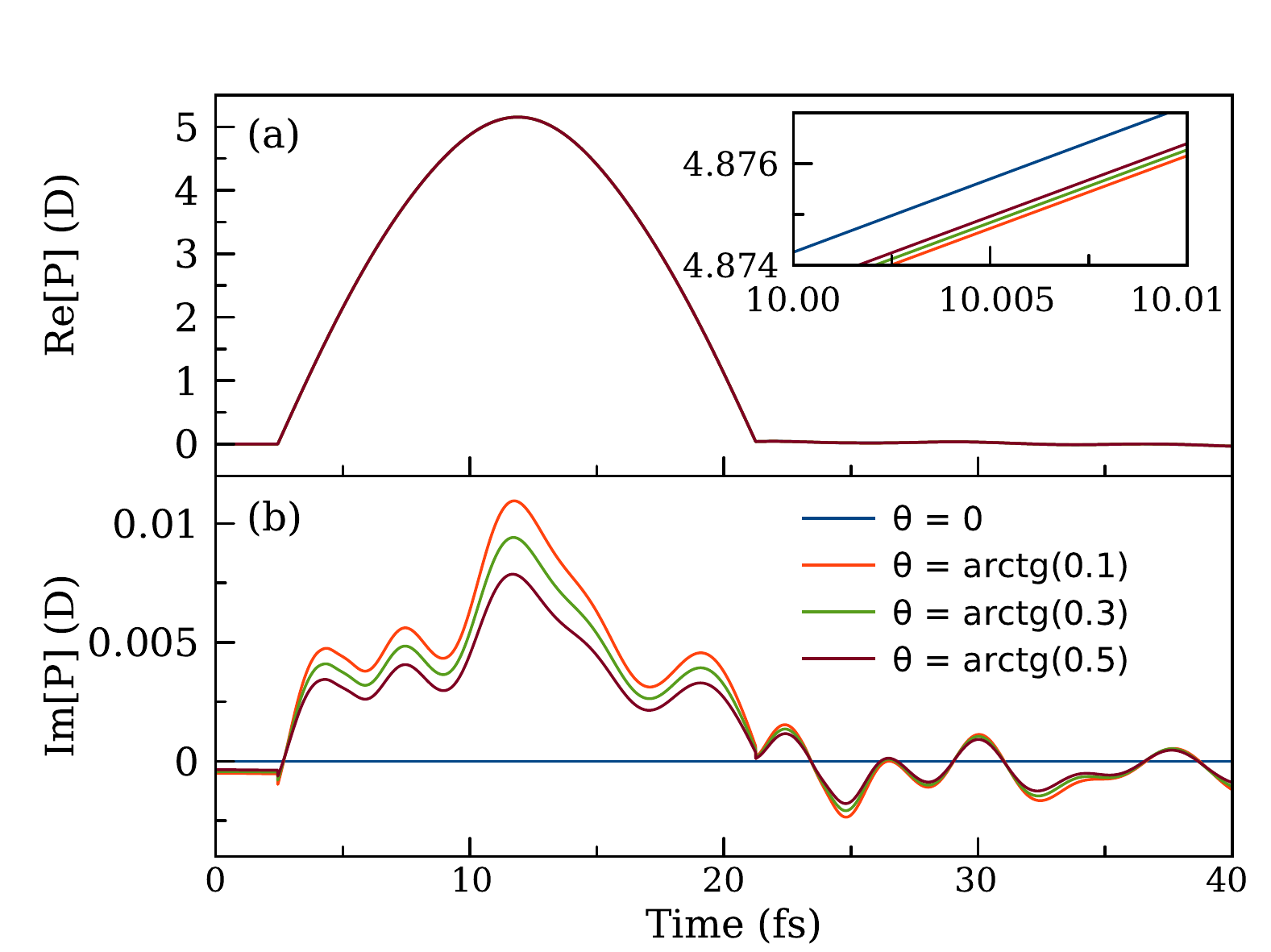}
\caption{The real (a) and imaginary (b) parts of the numerically calculated polarization for different angles $\theta$ of the complex rotation.}
\label{compareCS}
\end{center}
\end{figure}

\section{Results}

\subsection{Angle of the complex rotation}
In our calculations, we modeled an electron in the one-dimensional rectangular QW~(\ref{U0}) of the width $a=2.8$~nm and the barrier height $V=1$~eV. The system was excited by a pair of incident pulses~(\ref{Et}) with the delay of about $\tau_{d}=\pi/w_{21}=18.81$~fs. The field amplitude of the incident pulse was initially chosen as $E_{0}=3\times 10^{8}$~V/cm. The width of the Gaussian pulses was varied in the course of the study from a few tens to a few attoseconds.

Since our numerical method employs the complex-scaling technique, the obtained results become dependent on the rotation angle $\theta$ into the upper complex half-plane. In particular, the calculations of the dipole moment gain ambiguity because the integrated quantities, namely the wave function and the scaled coordinate variable in Eq.~(\ref{DM}), become dependent on the angle $\theta$ of the rotation.
However, a quality of the results should not depend on the artificially introduced parameters.
Therefore, first of all, we compared the numerically obtained polarization for different angles of the complex rotation with the same quantity without scaling.
In Fig.~\ref{compareCS} one can see the results of calculations of the polarization made for the Gaussian width $\tau_{p}$ of $2$~attosec. Indeed, the calculated quantity slightly depends on the complex rotation, specifically the polarization converges to the unscaled one as the angle of the rotation $\theta$ increases. Nevertheless, this dependence is of the order of the uncertainty of the numerical method and, thus, does not appreciably affect the quality of the results. As it had been shown earlier in Ref.~\cite{Moiseyev}, the complex scaling does not affect the square-integrable solutions, i.e. the bound states. Moreover, since we use the exterior complex-rotation, the function $R(x)$ equals to $x$ in the domain of the significantly nonzero support of the wave function. As a result, the calculated polarization does not change notably.
\begin{figure}[th!]
\begin{center}
\includegraphics[angle=0, width=0.5\linewidth]{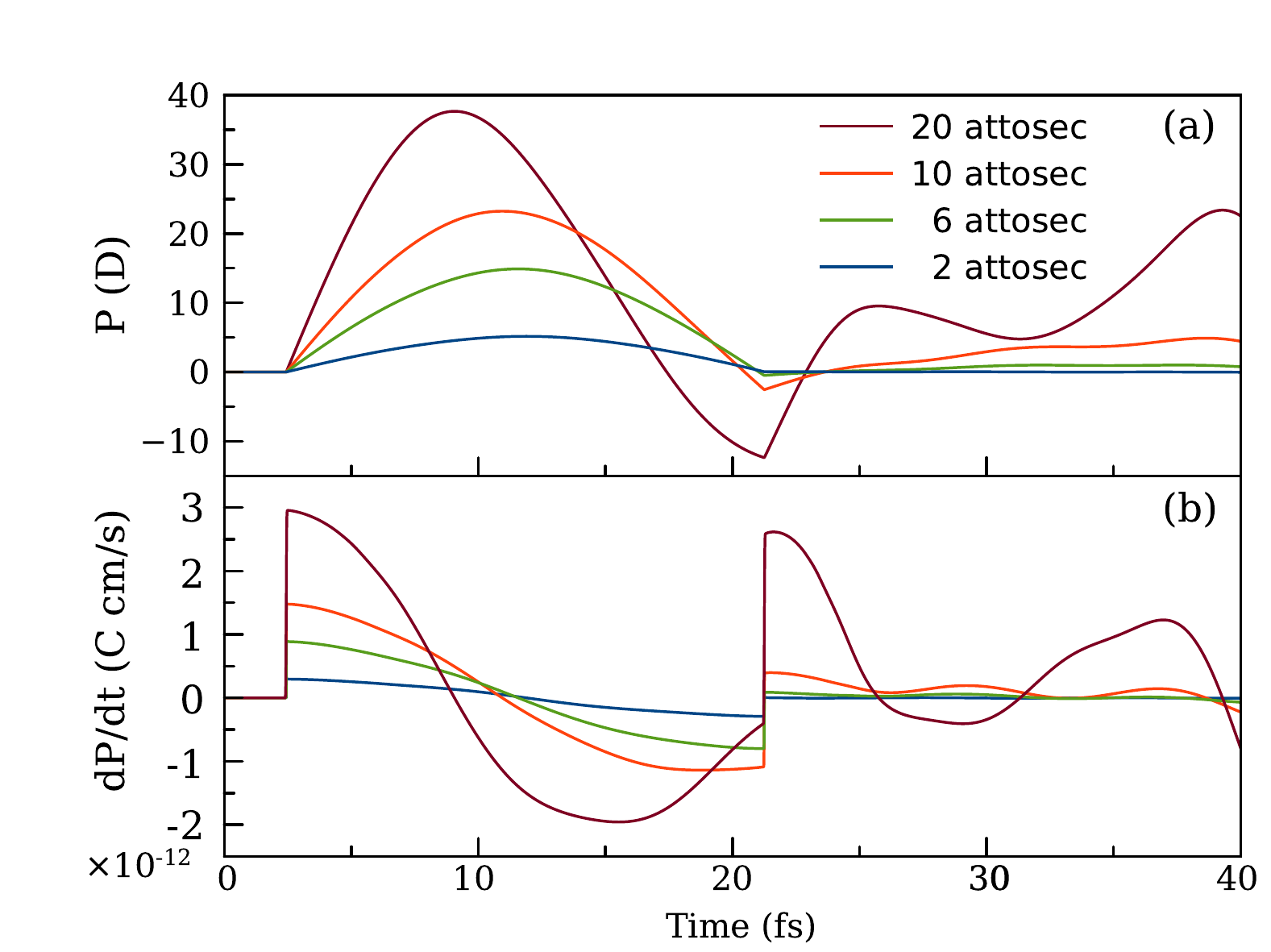}
\caption{Calculated polarization (a) and its first time derivative (b) resulted from the incident wave consisting of two Gaussian pump pulses in a rectangular QW with five bound states. The widths $\tau_{p}$ of the Gaussian pulses are $20$, $10$, $6$, and $2$ attoseconds.}
\label{compareP}
\end{center}
\end{figure}

\subsection{The stopped polarization pulse}
We performed calculations of the polarization for the rectangular QW having five bound states subjected to the incident Gaussian pulses of different widths $\tau_{p}$.
Other parameters of the system are as in the previous section.
The energy levels of the bound states are $E_{1}=0.037$~eV, $E_{2}=0.147$~eV, $E_{3}=0.328$~eV, $E_{4}=0.547$~eV, $E_{5}=0.865$~eV.
Fig.~\ref{compareP} shows the simulation results with a successive decrease of the widths of Gaussian pulses: $\tau_{p}=20,10,6,2$ attosec.
Our analysis shows that, as the width $\tau_{p}$ of the incident Gaussian pulses decreases to several attoseconds, the first pulse excites oscillations of the polarization, while the second one stops them. As a result, a polarization half-wave, or SPP, appears between the incident pulses. So, by decreasing the width of the pump pulses, it is possible to ensure that the second pulse suppresses polarization oscillations almost to zero.
Together with the polarization, one can see in the figure that the first derivative of the polarization, which is a solution of the wave equation for the superradiance pulse field~\cite{Morgner}, also vanishes after the second short pulse.

\begin{figure}[h]
\begin{center}
\begin{minipage}[t]{.45\textwidth}
  \centering
  \includegraphics[angle=0, width=1.0\linewidth]{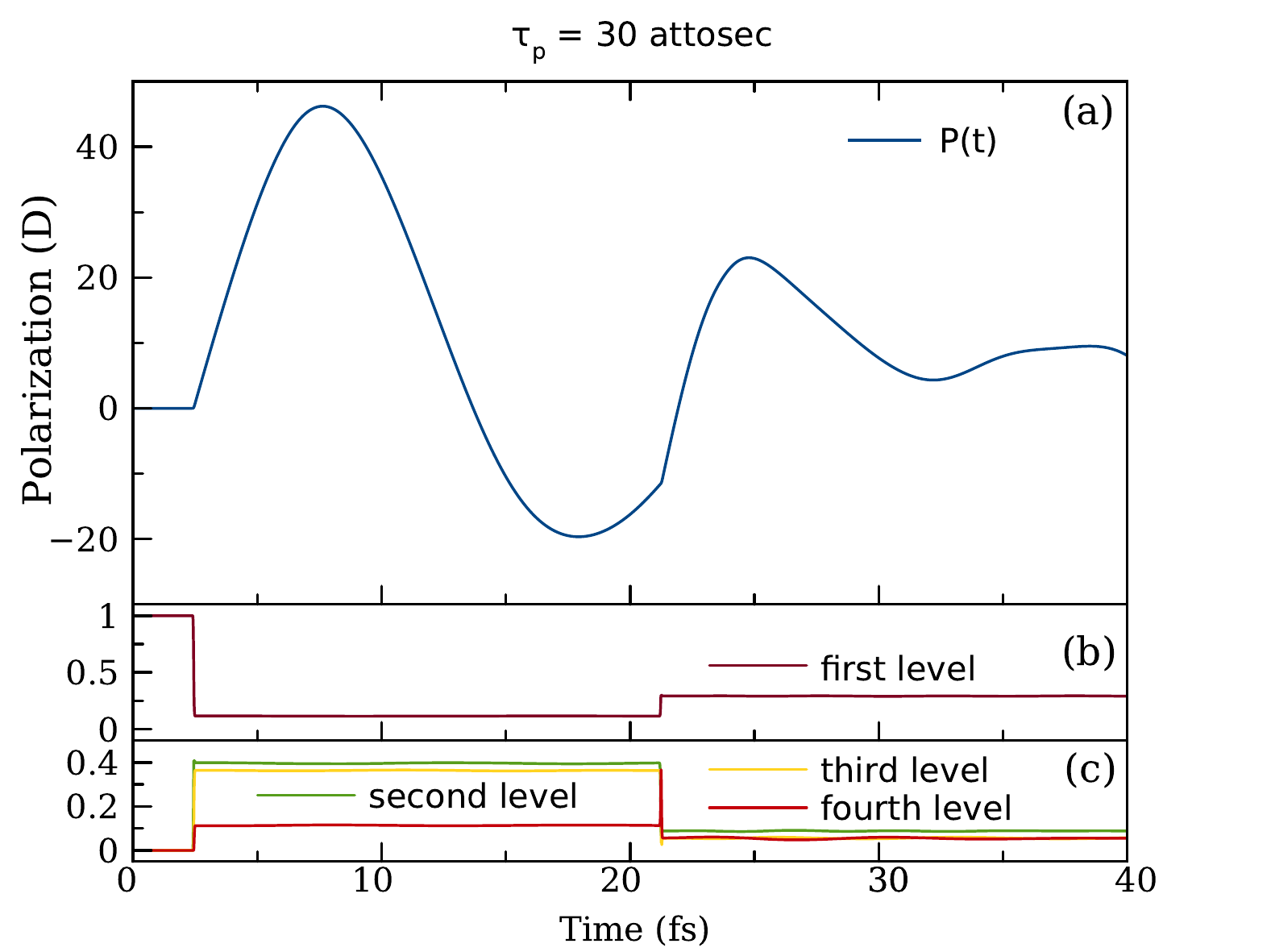}
\end{minipage}
\hfill
\begin{minipage}[t]{.45\textwidth}
  \centering
  \includegraphics[angle=0, width=1.0\linewidth]{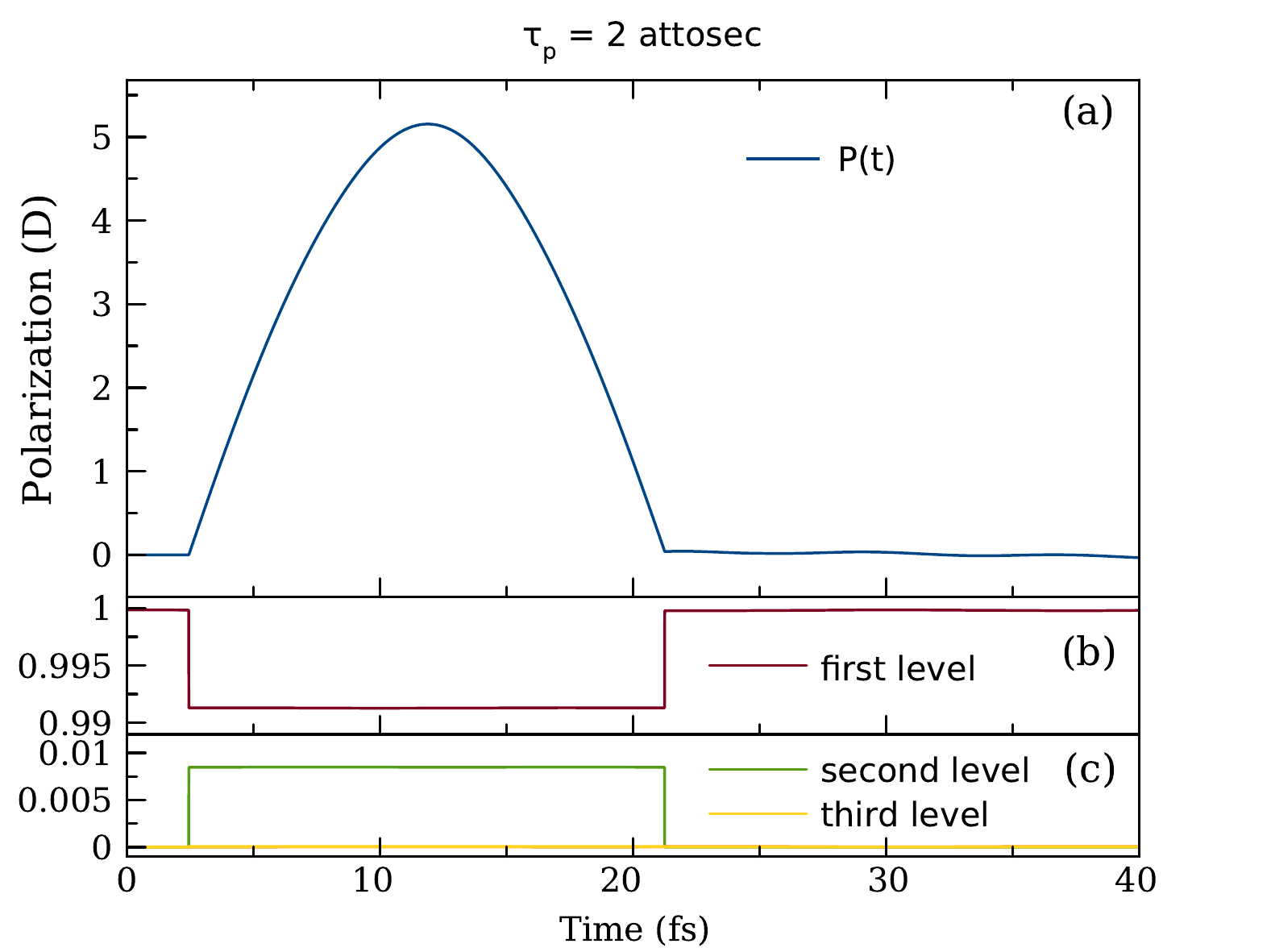}
\end{minipage}
\caption{The polarization (a) resulted from an incident wave consisting of two Gaussian pulses as well as populations of the lowest energy level (b) and excited levels (c) in the quantum well with five bound states. The results for the Gaussian pulses of width $\tau_{p}=30$~attosec \emph{(left)} and $\tau_{p}=2$~attosec \emph{(right)} are shown.}
\label{fig0230}
\end{center}
\end{figure}

In Fig.~\ref{fig0230}, the calculated polarizations as well as the populations of bound states are shown for $\tau_{p}=30$~attosec and $\tau_{p}=2$~attosec. It becomes clear that, as the width of the incident pulses decreases, the number of populated levels also decreases.
For $\tau_{p}=30$~attosec, after the first pulse, a noticeable simultaneous population of several lowest states occurs. Our study shows that for $\tau_{p}=10$~attosec and less, only a weak excitation of the second energy level takes place, and the higher levels remain practically unpopulated.
For small $\tau_{p}$, the dependencies of the populations on time do not qualitatively change. After the second Gaussian pulse, the population of the ground state returns to the initial unit level, and the populations of the remaining states again become equal to zero.
Correspondingly, Fig.~\ref{fig0230}, right plot, clearly shows the resulting polarization half-wave after the first pulse of width $\tau_{p}=2$~attosec and the complete absence of polarization after the second pulse.
For large values of $\tau_{p}$, several levels in the QW are populated at once, and after the second pulse they remain populated.
Indeed, Fig.~\ref{fig0230}, left plot, shows that after the first pulse of width $\tau_{p}=30$~attosec the polarization oscillates and the second pulse does not stop it. It is also worth noting that the obtained results remain valid for larger numbers of bound states in the QW.

\begin{figure}[h]
\begin{center}
\begin{minipage}[t]{.45\textwidth}
  \centering
  \includegraphics[angle=0, width=1.0\linewidth]{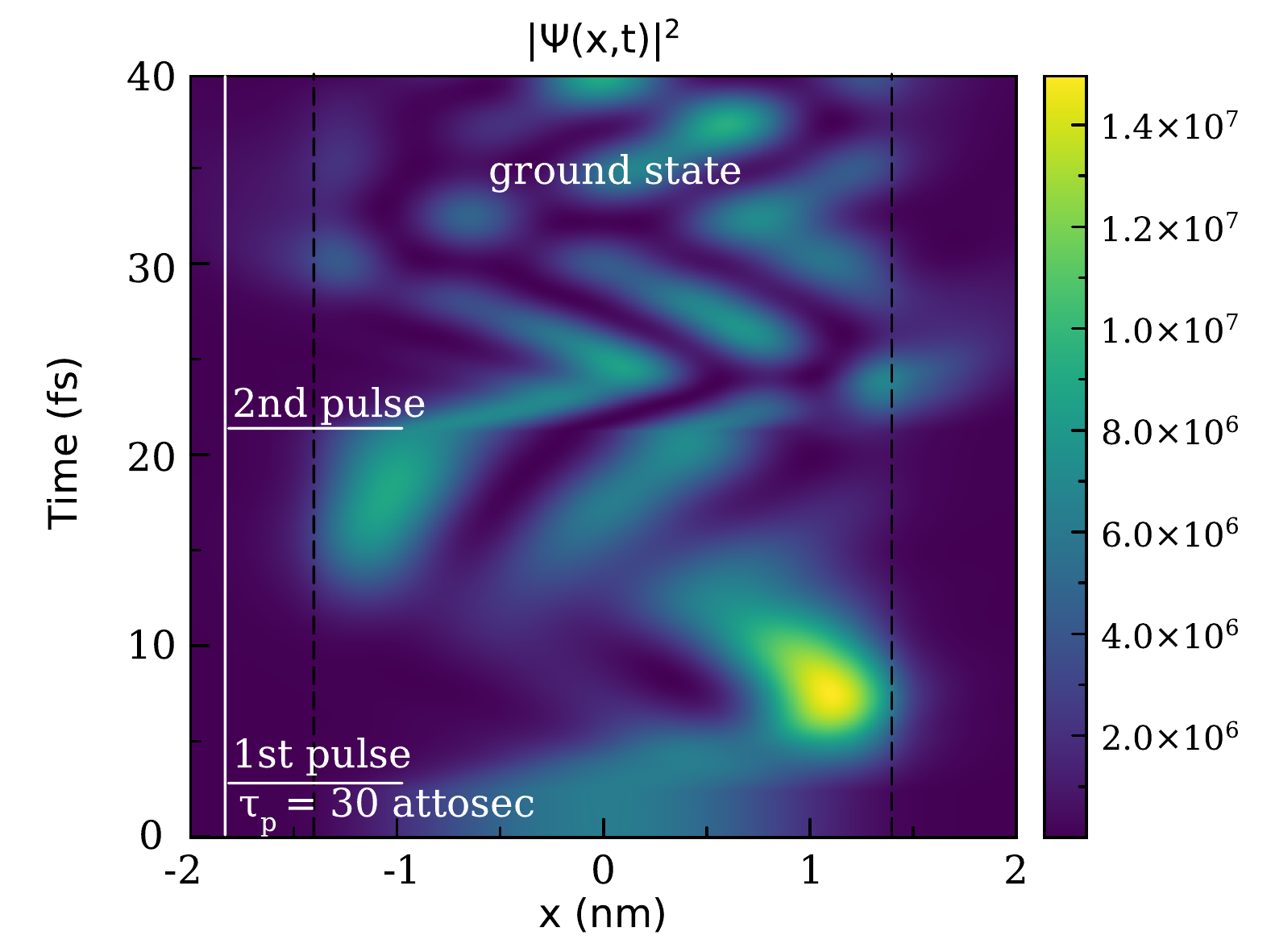}
\end{minipage}
\hfill
\begin{minipage}[t]{.45\textwidth}
  \centering
  \includegraphics[angle=0, width=1.0\linewidth]{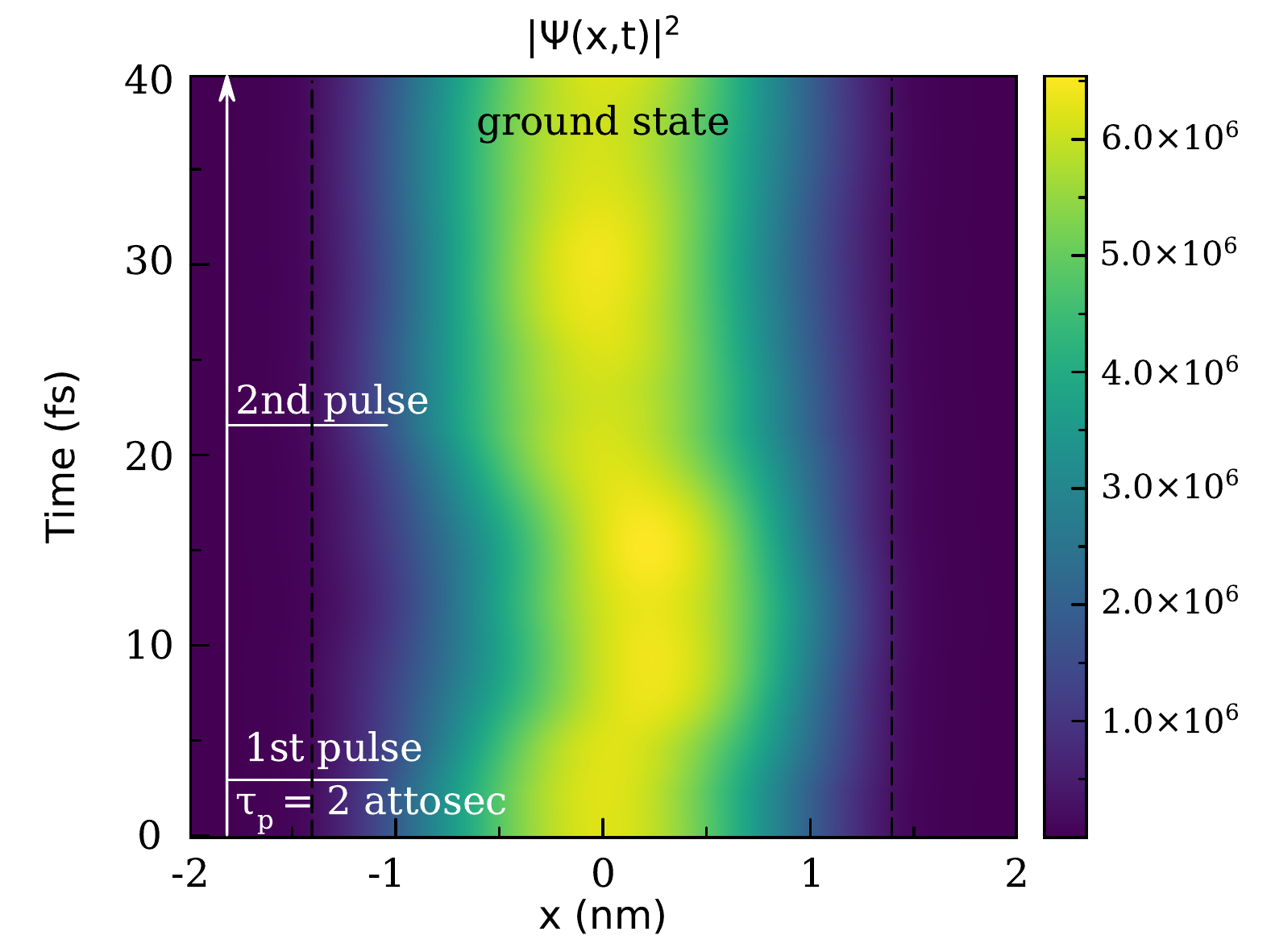}
\end{minipage}
\caption{The calculated ground state wave packet for the excitation of the system by two Gaussian pump pulses with the widths $\tau_{p}=30$~attosec \emph{(left)} and $\tau_{p}=2$~attosec \emph{(right)}.}
\label{packet302}
\end{center}
\end{figure}

The calculated wave packets are shown in Fig.~\ref{packet302} for the excitation by the wide pulses ($\tau_{p}=30$~attosec) and by the short pulses ($\tau_{p}=2$~attosec). The modeling shows that after the wide unipolar Gaussian pulses, the time-evolved wave packet is completely destroyed. As a result, it produces the irregular evolution of the polarization. On the other hand, the short pulse does not distort the wave packet. Instead, it makes the wave packet harmonically oscillate around the equilibrium in the center of the rectangular QW. Thus, the polarization also oscillates. The second short pulse almost completely stops the oscillations (the wave packet stops at the center of the QW) and the polarization becomes zero.

\subsection{Electric pulse area}
The dependence of populations on the width of the incident Gaussian pulses can be described by comparison of the electric pulse area with the characteristic atomic scale of the system~\cite{SE01,SE02}.
The former is a quantitative measure of the effect of the incident pulse on a system, namely, of a depopulation of the ground state and an excitation of the upper levels. The latter is the electric pulse area unit in terms of the characteristic parameters of the system. In Ref.~\cite{SE01} the characteristic atomic scale is defined by parameters of the electron in a QW as $S_{A}=2\hbar/ea$. For our system, $S_{A}=4.7\times 10^{-9}$~V$\,$sec/cm. For the Gaussian pulse, the electric pulse area is defined as $S_{E}=E_{0}\sqrt{\pi}\tau_{p}$. Since $E_{0}=3\times 10^{8}$~V/cm, it gives $S_{E}=1.06 \times 10^{-9}$~V$\,$sec/cm for $\tau_{p}=2$~attosec and $S_{E}=16 \times 10^{-9}$~V$\,$sec/cm for $\tau_{p}=30$~attosec.
So, for short pulses $S_{E} \ll S_{A}$ and for wide ones $S_{E} \gg S_{A}$.
Based on these estimations, we can conclude that the ultrashort incident pulse does not affect the ground state population and just slightly populates the excited ones, mainly the first excited one. Therefore, we obtain the polarization oscillations with a frequency defined by the main resonant transition $w_{21}=(E_{2}-E_{1})/\hbar$.

\begin{figure}[h!]
\begin{center}
\includegraphics[angle=0, width=0.5\linewidth]{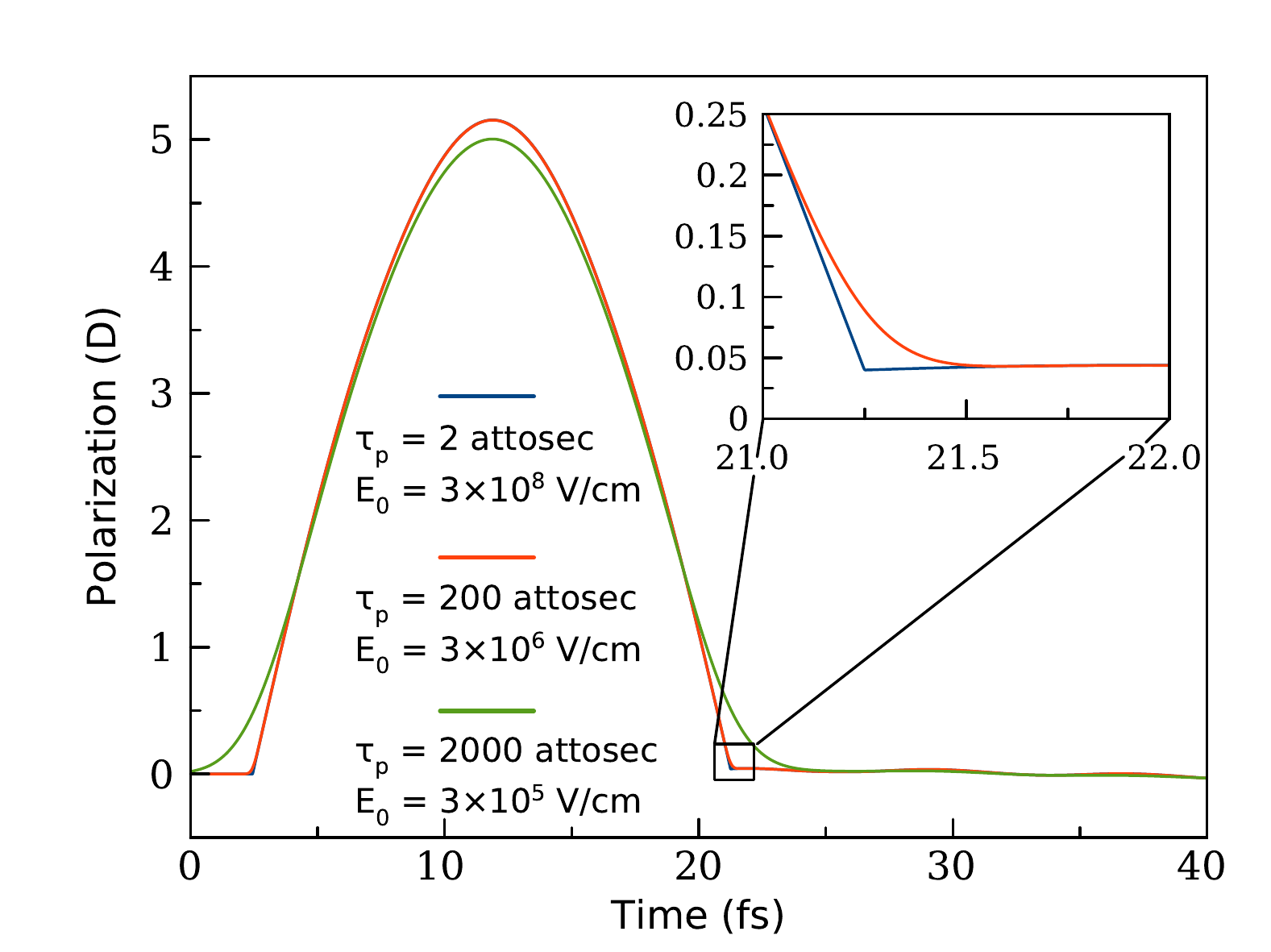}
\caption{The polarizations obtained from the incident Gaussian pulses defined by different widths and amplitudes. The width $\tau_{p}$ and amplitude $E_{0}$ of the incident pulses were simultaneously varied in such a way as to keep the electric pulse area $S_{E}=E_{0}\sqrt{\pi}\tau_{p}$ to be constant.}
\label{figEPAsame}
\end{center}
\end{figure}

One can see that the electric pulse area $S_{E}$ of the Gaussian pulse depends only on the pulse amplitude $E_{0}$ and the width $\tau_{p}$.
Therefore, the effect of the incident pulse on the system does not change, if $E_{0}$ and $\tau_{p}$ are simultaneously altered in a way to keep $S_{E}$ to be the same. This fact is shown in Fig.~\ref{figEPAsame}. There are the polarizations obtained for the initial parameters $E_{0}=3\times 10^{8}$~V/cm and $\tau_{p}=2$~attosec as well as for alternative values: $E_{0}=3\times 10^{6}$~V/cm and $\tau_{p}=200$~attosec. One can see that since the electric pulse area is the same for such pulse parameters, the polarizations are also almost identical. They only slightly change in the domain where the incident short pulses are nonzero, see the inset in the figure.
The polarization resulted from the pulse with $E_{0}=3\times 10^{5}$~V/cm and $\tau_{p}=2000$~attosec is also shown for comparison. One can see that such a pulse width, $\tau_{p}$, cannot be considered as to be much smaller than the half-period, $\pi/w_{21}$, of the main resonant transition oscillation. Thus, the polarization more noticeably deviates from the one obtained for $\tau_{p}=2$~attosec.

\section{Conclusion}
We showed that during an excitation of the five-level rectangular QW by the pair of short unipolar pulses the polarization oscillates with a frequency $w_{21}$ determined by the main resonant transition.
It means that in such a case only the first and the second energy levels become populated.
The second pulse launched after the half-period of this oscillation, $\pi/w_{21}$, almost completely stops the oscillations.
As a result, the SPP can be obtained under the short unipolar pulse excitation not only in the two-level systems, but also in the multi-level setup.
For wide incident pulses, the upper energy levels become populated and the SPP does not appear.
A comparison of the electric pulse area and the characteristic atomic scale of the system allowed us to quantitatively estimate the effect of the incident pulse.

\section*{Acknowledgments}

The authors are grateful to the Russian Science Foundation, grant no. 21-72-10028, for the financial support.
The calculations were made using the facilities of the ``Computational center of SPbU''.

\begin{figure}[th!]
\begin{center}
\includegraphics[angle=0, width=0.5\linewidth]{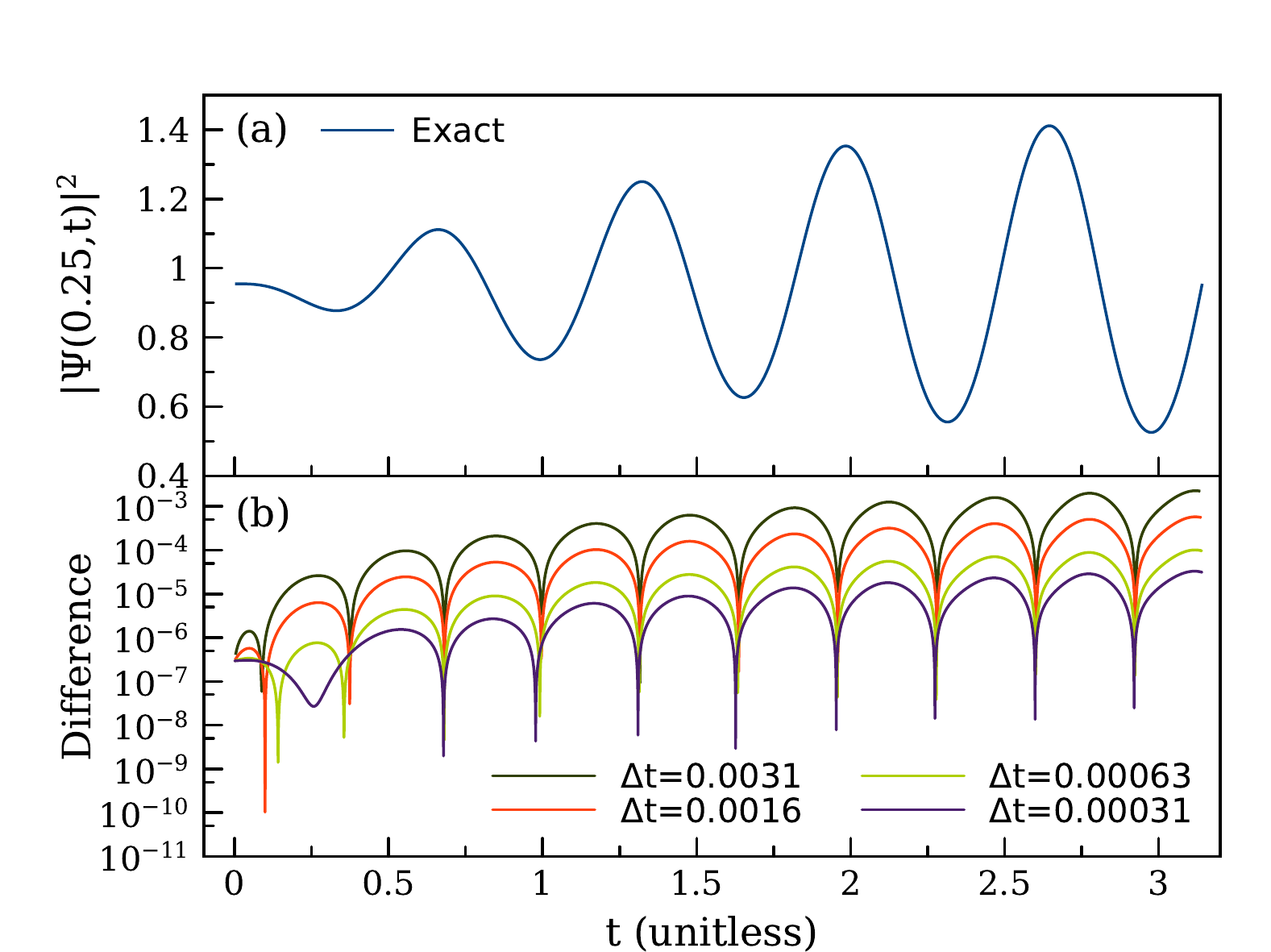}
\caption{ (a) The time evolution of the slice ($x=0.25$) of the quantum oscillator ground state wave function squared under $\sin{(w_{0}t)}$ excitation. (b) The difference between exact and numerical solutions for $|\Psi(0.25,t)|^{2}$ as a function of $t$.}
\label{uAB}
\end{center}
\end{figure}

\section*{Appendix}
The numerical scheme for solving Eq.~(\ref{SE}) with the complex scaling (\ref{Rx}) is following.
Let us introduce the equidistant knots over spatial and time variables as $x_{j}$, $j=1,\ldots,J$ and $t_{n}$, $n=1,\ldots,N$, respectively. The grid steps are denoted as $\Delta x=x_{j+1}-x_{j}$ and $\Delta t=t_{n+1}-t_{n}$.
Then, for the grid function $\Psi_{j}^{n}=\Psi(x_{j},t_{n})$ the Crank-Nicolson numerical scheme~\cite{CN} used in our work reads as
\begin{equation}
\frac{\Psi_{j}^{n+1}-\Psi_{j}^{n}}{\Delta t} = \frac{i \hbar}{2m} \frac{1}{(R'_{x})^2} \delta^2 \Psi_{j}^{n+1/2} - \frac{i \hbar}{2m} \frac{R''_{xx}}{(R'_{x})^{3}} \delta \Psi_{j}^{n+1/2} - \frac{i }{\hbar} \left( \frac{V_{j}^{n+1}+V_{j}^{n}}{2} \right) \Psi_{j}^{n+1/2}, 
\end{equation}
where $V^{n}_{j}=V(x_{j},t_{n})$ and $\Psi_{j}^{n+1/2}$ denotes the averaged sum $(\Psi_{j}^{n+1}+\Psi_{j}^{n})/2$. The finite-difference operators act on functions in the averaged sum as
$$
\delta \Psi_{j}^{n} = \frac{\Psi_{j+1}^{n}-\Psi_{j-1}^{n}}{2\Delta x}, \quad
\delta^{2} \Psi_{j}^{n} = \frac{\Psi_{j+1}^{n}-2 \Psi_{j}^{n}+ \Psi_{j-1}^{n}}{(\Delta x)^{2}}.
$$
This scheme leads to the iterative solution over time with solving the linear equation at each fixed time knot.
The matrices of this equation are block-tridiagonal~\cite{Korneev,Yak,Nugumanov} and, thus, allow relatively fast solution at each step and, as a result, fast modeling of the time evolution.

It is worth noting that the Crank-Nicolson numerical scheme calculates a solution with the second order uncertainty both in space and time: $\sim \left((\Delta x)^{2}+(\Delta t)^{2}\right)$~\cite{CNconvergence1,CNconvergence2}.
We verified the uncertainty of the numerical scheme by comparing the calculated wave function (without complex scaling) with the exact analytical solution for the time evolution of the ground state wave function of a particle confined in the parabolic QW, $V(x)=mw^{2}\,x^{2}/2$, where $w$ is a fixed parameter. To achieve a nontrivial time-dependence of the solution, this system was subject to the incident sin-like excitation $E(t)=\sin{(w_{0}t)}$~\cite{Baz,Hamprecht,AmJPhys} with $w_{0} \ne w$.
This incident wave alters the standard $\sim \exp{(iwt)}$ time-evolution
of the stationary wave-packet of the ground state
$$
\Psi(x,0) = \left(\frac{m w}{\hbar \pi }\right)^{1/4} \exp{\left(-\frac{mw}{\hbar} \frac{x^{2}}{2}\right)},
$$
which was chosen as the initial function for the Cauchy initial value problem over time.

Fig.~\ref{uAB} shows the time evolution of the slice of the exact quantum oscillator ground state wave function squared under the sin-like excitation as well as the difference between the exact and our numerical solutions. In calculations, the material parameters are taken to be $\hbar=e=m=1$, $w=10$, $w_{0}=9$. The computed solutions with different grid steps allowed us to obtain the convergence rate of the numerical scheme which coincided with the theoretical one.

\section*{References}


\end{document}